\documentclass[11pt]{article}
\pdfoutput=1
\usepackage[margin=1.0in]{geometry}
\usepackage{amsmath,amssymb,graphicx}
\usepackage{hyperref}
\usepackage{slashed}

\usepackage[utf8]{inputenc}

\usepackage{upgreek}

\usepackage[font=small,labelfont=bf]{caption}
\usepackage{cite}

\newcommand{\rmd}{{\mathrm d}}
\newcommand{\labD}{{\scalebox{0.65}{$\Delta$}}}
\DeclareMathOperator{\tr}{tr}

\newcommand{\eB}{e_B}

\newcommand{\be}{\begin{equation}}
\newcommand{\ee}{\end{equation}}

\usepackage{xcolor}
\usepackage{bm}

\title{\bf  The Weak Gravity Conjecture and Axion Strings}
\author{Ben Heidenreich${}^1$,  Matthew Reece${}^2$, and Tom Rudelius${}^3$ \\
{\small ${}^1$Department of Physics, University of Massachusetts, Amherst, MA 01003 USA}\\
{\small ${}^2$Department of Physics, Harvard University, Cambridge, MA, 02138, USA}\\
{\small ${}^3$Department of Physics, University of California, Berkeley, CA 94720, USA}}

\begin{document}

\begingroup
{\flushright ACFI-T21-10\par}
\let\newpage\relax%
\maketitle
\endgroup

\begin{abstract}
Strong (sublattice or tower) formulations of the Weak Gravity Conjecture (WGC) imply that, if a weakly coupled gauge theory exists, a tower of charged particles drives the theory to strong coupling at an ultraviolet scale well below the Planck scale. This tower can consist of low-spin states, as in Kaluza-Klein theory, or high-spin states, as with weakly-coupled strings. We provide a suggestive bottom-up argument based on the mild $p$-form WGC that, for any gauge theory coupled to a fundamental axion through a $\theta F \wedge F$ term, the tower is a stringy one. The charge-carrying string states at or below the WGC scale $g M_\mathrm{Pl}$ are simply axion strings for $\theta$, with charged modes arising from anomaly inflow. Kaluza-Klein theories evade this conclusion and postpone the appearance of high-spin states to higher energies because they lack a $\theta F \wedge F$ term. For abelian Kaluza-Klein theories, modified arguments based on additional abelian groups that interact with the Kaluza-Klein gauge group sometimes pinpoint a mass scale for charged strings. These arguments reinforce the Emergent String and Distant Axionic String Conjectures. We emphasize the unproven assumptions and weak points of the arguments, which provide interesting targets for further work. In particular, a sharp characterization of when gauge fields admit $\theta F \wedge F$ couplings and when they do not would be immensely useful for particle phenomenology and for clarifying the implications of the Weak Gravity Conjecture.

\end{abstract}

\section{Introduction}

The Weak Gravity Conjecture (WGC) postulates that, when a consistent theory of quantum gravity includes a gauge group with coupling $g$, there is a particle carrying gauge charge $q$ with a mass below the scale $q g M_\mathrm{Pl}$~\cite{Arkanihamed:2006dz}. More recently, it has been appreciated that a stronger statement holds in all well-understood theories: an infinite {\em tower} of charged particles exists, beginning at or below the scale $g M_\mathrm{Pl}$. This condition has been formalized in the Sublattice Weak Gravity Conjecture (sLWGC)~\cite{Heidenreich:2015nta,Heidenreich:2016aqi, Montero:2016tif} and the Tower Weak Gravity Conjecture (TWGC)~\cite{Andriolo:2018lvp, Heidenreich:2019zkl}. We will refer to the original WGC as the ``mild WGC,'' to distinguish it from the stronger T/sLWGC.\footnote{We will assume that the mild  WGC holds  for $q$ an $O(1)$ number, and similarly that the sublattice index in the sLWGC case or sparseness in the TWGC case are $O(1)$. These assumptions hold in all known examples, though they do go beyond the strict statements of the conjectures.} The towers of charged particles predicted by the T/sLWGC can take on distinct forms in different theories. In Kaluza-Klein theories, they are simply KK modes of the light particles in the theory. Local quantum field theory continues to apply well above the scale $g M_\mathrm{Pl}$ when $g$ is small; it just becomes local QFT in a larger number of dimensions. Local QFT must break down by the higher-dimensional Planck scale, but this can scale as $g^r M_\mathrm{Pl}$ where $r < 1$. By contrast, in some cases the tower of states consists of excitations of a low-tension string. The charged modes include high-spin states, and local QFT with a finite number of fields is no longer valid. Indeed, it has been proposed that these two options---the presence of extra dimensions as in Kaluza-Klein theory, or of low-tension strings---are the only possibilities for realizing a small-$g$ gauge theory in quantum gravity~\cite{Lee:2019wij}, a claim (called the Emergent String Conjecture) supported by many studies of well-understood corners of the string landscape~\cite{Grimm:2018ohb,Lee:2018urn,Grimm:2018cpv,Corvilain:2018lgw,Lee:2019tst,Font:2019cxq,Lee:2019xtm,Baume:2019sry,Gendler:2020dfp,Xu:2020nlh,Lanza:2020qmt,Heidenreich:2020ptx,Lanza:2021qsu,Palti:2021ubp}. (These studies relate the WGC to the more general  Distance Conjecture regarding infinite-distance limits in quantum gravity~\cite{Ooguri:2006in,Klaewer:2016kiy,Heidenreich:2018kpg,Stout:2021ubb}.)

It is important, both for conceptual questions about quantum gravity and for possible phenomenological applications, to know which type of tower of charged states (and related ultraviolet cutoff) arises in a given application. The purpose of this note is to provide a simple, bottom-up argument based on the mild WGC that in a large class of 4d theories containing gauge fields coupled to a light axion field $\theta$ through a Chern-Simons coupling $\theta F \wedge F$, the WGC tower is a stringy one. In particular, there is a tower of charged modes that arise as excitations of an axion string. By axion string, we mean an object around which the periodic variable $\theta$ has nontrivial winding.  The argument that we present relies on assumptions that do not apply universally. In particular, it does not apply to Kaluza-Klein theory. We will review how Kaluza-Klein theories fail to satisfy the premises of the argument and show that in some cases a modified argument can still identify a stringy scale in Kaluza-Klein theory.

This argument fits nicely not only with the Emergent String Conjecture of~\cite{Lee:2019wij}, but also with the Distant Axionic String Conjecture of~\cite{Lanza:2020qmt,Lanza:2021qsu}. The latter postulates that infinite distance limits in 4d ${\cal N} = 1$ supersymmetric theories of quantum gravity can be viewed as RG flows into the core of an axion string. Our arguments do not assume supersymmetry, but they do assume that axionic couplings exist. To understand the relationship between the earlier conjectures, notice that the Emergent String Conjecture focuses on the tower of states that becomes light most quickly, which may be a Kaluza-Klein tower even when there is also a string becoming tensionless (at a slower rate). In such cases,~\cite{Lanza:2021qsu} conjectured that the mass scale of the tower always relates to the axion string tension as $m^2 \propto T^w$, with $w \in \mathbb{Z}_{> 0}$. We will discuss such an example below.

Before discussing the argument, we will review some useful concepts, particularly in cases with recently-introduced terminology that may not be familiar to readers. We will make use of the distinction between a {\em fundamental axion} and an ordinary pseudo-Goldstone boson, in language introduced in~\cite{Reece:2018zvv}. For an ordinary pseudo-Goldstone boson, axion strings are well-understood semiclassical objects. The core of the string is described within the low-energy EFT. For instance, $\theta$ may be the phase of a complex field $\Phi(x) = \frac{1}{\sqrt{2}} r(x) \mathrm{e}^{i \theta(x)}$, and if $\Phi(x)$ has a canonical kinetic term, then the point $r(x) = 0$ lies at finite distance in field space and is realized in the string core. By contrast, for a fundamental axion, the analogous point lies at infinite distance. This occurs, for instance, in some string-theoretic examples where the axion is the imaginary part of a complex modulus field $T(x) = t(x) + i \theta(x)$ and the kinetic term takes the form $\frac{1}{(T + T^\dagger)^2} \partial_\mu T^\dagger \partial^\mu T$. In this case, the core of the string probes a region at infinite distance in field space. It cannot be described purely within the low-energy EFT, and requires an ultraviolet completion of the theory. One source of such fundamental axions is from fundamental, higher-dimensional $p$-form gauge fields ($p \geq 1$), with $\theta  = \int_{\Sigma_p} C_p$ arising from the integral of the gauge field over a $p$-cycle. In this case, the $\theta F \wedge F$ coupling in four dimensions can originate from the dimensional reduction of a Chern-Simons term $C_p \wedge F \wedge F$ in the higher-dimensional theory. In examples of this type, the axion strings turn out to be fundamental objects like the F-strings or D-strings of string theory.  The arguments that we discuss in this paper, at least in their strongest form, will apply to gauge theories interacting with fundamental axions.

Another useful concept is that of a {\em Chern-Weil symmetry}~\cite{Heidenreich:2020pkc}. Chern-Weil global symmetries are a particular class of global symmetries in quantum field theory whose Noether currents are given by wedge products of gauge field strengths, such as $F \wedge H$ and $\tr(F \wedge F)$, and their conservation ${\rm d} J = 0$ follows from Bianchi identities. In quantum gravity, it has been argued convincingly that exact global symmetries do not exist~\cite{Banks:2010zn, Harlow:2018tng}, which means that these Chern-Weil global symmetries must be either gauged or broken when the quantum field theory is coupled to gravity. Many familiar phenomena in string theory can be understood as consequences of this gauging/breaking ~\cite{Heidenreich:2020pkc}.

A Noether current of degree $k$ generates a $(d-k-1)$-form global symmetry in $d$ dimensions, so in particular a 4-form current like $\tr(F \wedge F)$ or $F \wedge H$ generates a $(d-5)$-form Chern-Weil global symmetry. Such a symmetry is clearly sensible in $d \geq 5$ dimensions, but in $d=4$ dimensions the current generates a ``$(-1)$-form global symmetry''  \cite{Cordova:2019jnf, Tanizaki:2019rbk, McNamara:2020uza, Heidenreich:2020pkc}. At face value, ``$(-1)$-form symmetry'' is ill-defined, as a 4-form current $J$ is trivially conserved in 4d. However, there is a sense in which such a symmetry exists and defines a conserved charge $q = \oint_{\text{spacetime}} J$, which is a topological invariant of the theory~\cite{Heidenreich:2020pkc}.\footnote{For $J=\tr(F \wedge F)$, this topological invariant is simply instanton number.} In quantum gravity, we expect that a $(-1)$-form global symmetry should be broken or gauged, the latter of which is accomplished by coupling the Noether current to a 0-form gauge field, namely, an axion $\theta$. It is not surprising, therefore, that axion couplings such as $\theta \tr(F \wedge F)$ and $\theta F \wedge H$ are ubiquitous in known 4d string vacua: such couplings play the necessary role of eliminating $(-1)$-form Chern-Weil global symmetries by gauging them. And as noted above and explained in more detail below, these couplings are closely tied to our key question of when the WGC tower is a stringy one.

In \S\ref{sec:argument}, we present our argument and emphasize the underlying assumptions and their possible weak points. Our argument relies crucially on anomaly inflow for axion strings; for the convenience of readers unfamiliar with the details, we provide a review in Appendix~\ref{app:inflow}. In \S\ref{sec:KKtheory}, we focus on Kaluza-Klein theory as a source of examples that do not satisfy the assumptions, and explain how additional ingredients (supersymmetry or winding number symmetry) allow the argument, in a suitably modified form, to nonetheless be applied. \S\ref{sec:5dsugra} explains how 5d supergravity theories exemplify our arguments. Finally, \S\ref{sec:conclude} offers some remarks on directions for future research.

\section{The Argument and Its Premises}
\label{sec:argument}

We consider gauge theories with a $\theta F \wedge F$ coupling (in the abelian case) or $\theta \tr(F \wedge F)$ (in the nonabelian case). We keep track of factors of $2\pi$ carefully, but not of other order-one factors appearing in the WGC (which can depend on details of couplings to scalar moduli fields). We consider the axion to have periodicity $\theta \cong \theta + 2\pi$, and take the action
\begin{equation}
S = \int \Big[ -\frac{1}{2 g^2} F \wedge \star F - \frac{1}{2} f_\theta^2 \rmd \theta \wedge \star \rmd \theta + \frac{1}{8\pi^2} \theta F \wedge F \Big].
\end{equation}
In the nonabelian case, we should send $F \wedge \star F \to 2 \tr(F \wedge \star F)$, and $F \wedge F \to \tr(F \wedge F)$.

As a first step, we consider the axion (0-form) WGC, which tells us that the axion couples to instantons with action bounded by
\begin{equation} \label{eq:axionWGC}
S_\mathrm{inst} \lesssim \frac{M_\mathrm{Pl}}{f_\theta}.
\end{equation}
The usual BPST gauge theory instantons, for non-abelian gauge groups, have action $S_\mathrm{inst} = 8\pi^2/g^2$. Remarkably, the same scaling holds for abelian instantons arising from monopole loops~\cite{Fan:2021ntg}, as a consequence of the Witten effect~\cite{Witten:1979ey}.\footnote{This statement holds only when the monopole core radius is sufficiently large relative to the scale of axion screening near the monopole; however, we show in Appendix~\ref{app:monopolecheck} that our conclusions hold in the opposite limit as well, this time using the mild form of the WGC for magnetic monopoles.} We now \emph{assume} that the instantons that obey the bound~\eqref{eq:axionWGC} are such gauge theory instantons, and comment below on why this is often a reasonable assumption. This assumption tells us that
\begin{equation}
f_\theta \lesssim \frac{g^2}{8\pi^2} M_\mathrm{Pl}.
\end{equation}
This is already an interesting statement. Using the GUT coupling estimate $\alpha_\mathrm{GUT} = g^2/(4\pi) \approx 1/25$, it gives $f_\theta \lesssim 1.5 \times 10^{16}\,\mathrm{GeV}$ for an axion coupling to the Standard Model gauge groups. This claim is interesting, independent of the rest of the argument. It has been observed before that the axion WGC could lead to such a conclusion, but some of the premises now appear stronger in light of the viewpoint that axions exist to avoid unbroken global Chern-Weil symmetries.

In the next step, we apply the {\em magnetic} axion WGC, i.e., we apply the WGC to the 2-form $B$-field dual to $\theta$. This field has a dimensionful coupling constant $\eB = 2 \pi f_\theta$, with the relative factor of $2\pi$ being the standard  one from Dirac quantization. The WGC applied to the $B$-field implies the existence of a string with tension obeying the inequality
\begin{equation} \label{eq:stringtensionbound}
T \lesssim \eB M_\mathrm{Pl} \lesssim \frac{g^2}{4\pi} M_\mathrm{Pl}^2.
\end{equation}
Recalling that the tension of a string scales as $T = \frac{1}{2\pi} M_\mathrm{str}^2$, this translates into the mass scale for string excitations:
\begin{equation} \label{eq:stringscale}
M_\mathrm{str} = \sqrt{2 \pi T} \lesssim g M_\mathrm{Pl}.
\end{equation}
In this way, we recover the WGC scale for our original gauge field solely from the WGCs for $\theta$ and $B$, together with the assumed form of the instanton action. (In particular, all of the $2\pi$ factors have canceled.)

How does this relate to the WGC itself? The string necessarily admits {\em charged} zero mode excitations, due to anomaly inflow~\cite{Callan:1984sa}. For readers who are unfamiliar with this argument, we have presented a self-contained review in Appendix~\ref{app:inflow}. The existence of charged modes on the string means that there are charged particles at the scale $M_\mathrm{str}$, in the form of closed loops of string with circulating charge. By~\eqref{eq:stringscale}, these charged particles satisfy the mild WGC (up to an $O(1)$ prefactor). Hence, the mild  WGC for our 1-form gauge field follows from the mild WGCs for 0-form and 2-form gauge fields and our assumption about the instanton action.

If we assume that $\theta$ is a fundamental axion, we can make a much stronger statement. 
In particular, the axion strings are now fundamental objects, i.e., the string core probes ultraviolet physics. 
The string therefore admits a tower of excitations involving new fundamental degrees of freedom. Among these excitations are those involving many quanta of the charged worldsheet modes circulating around a closed loop of string. These yield a tower of charged particle resonances with increasing mass and charge, from which we recover not only the mild WGC but a version of the T/sLWGC as well! (Likewise, the excitations include arbitrarily high spins, and signal the breakdown of local QFT near the string scale.)

How does this differ from non-fundamental axion strings associated to ordinary pseudo-Goldstone bosons? The latter have finite size related to the symmetry breaking scale and have ordinary local physics in their core. For such pseudo-Goldstone strings, the charged excitations living along the string do not lead to new degrees of freedom relative to those that are visible in the bulk; instead, the anomaly inflow argument merely allows us to deduce the presence of WGC-satisfying bulk charged fields from which these excitations arise.

Thus, the consequences of our argument are far more profound when $\theta$ is a fundamental axion. 
 If so there must be fundamental stringlike objects at or below the WGC scale, providing a concrete realization of the tower promised by the T/sLWGC (up to undetermined $O(1)$ prefactors), and signaling the transition from local QFT to quantum gravity.

\subsection{Summary of Assumptions}\label{ssec:Assumptions}

It is useful to highlight the various assumptions that have gone into this argument. Some of them are expected to hold quite generally in theories of quantum gravity, while others may be more limited. We attempt to order them from what we perceive to be the least general to the most general:

\begin{enumerate}

\item {\em An axion interacting with the gauge fields via a $\theta F \wedge F$ term.} Such an interaction is not always present. However, it is very common, for example, for gauge theories realized in intersecting D-brane models~\cite{Blumenhagen:2005mu}. As we discuss below, in some cases where it does not hold, a related statement nonetheless holds and allows a similar argument. In any case, for certain phenomenological applications one might wish to focus on gauge groups with such a coupling.

\item \label{number2} {\em Gauge-theory instanton dominance.} We have assumed that the instantons that should obey the axion WGC are gauge theory instantons, so that we can use the known formula for their action, $S_\mathrm{inst} = 8\pi^2/g^2$. In general, there could be {\em other} instantons with smaller action coupled to the same axion, which would weaken our argument. In QFT, there would be no restriction on this possibility. However, if we have two qualitatively different kinds of instantons coupled to the same axion, the axion only gauges one linear combination of two $(-1)$-form symmetries (counting the independent instanton numbers). In quantum gravity, we expect that the other linear combination must be either gauged, in which case there is a second axion and we should modify the argument, or explicitly broken. Breaking of the two $(-1)$-form symmetries to the diagonal entails a continuous process of deforming one kind of instanton into the other. In known examples, such as GUTs or the connection between small instantons and D-branes, we expect the actions of the different instantons to be related to each other. Thus, the Chern-Weil perspective of~\cite{Heidenreich:2020pkc} makes this assumption much more plausible than it might otherwise have been. The existence of instantons for abelian gauge theory, in the form of monopole loops with dyonic winding~\cite{Fan:2021ntg}, with parametrically the same instanton action also makes this step more general than one might have suspected.

The computation of $S_\mathrm{inst}$ depends on the scale at which the running coupling $g$ is evaluated. In asymptotically free gauge theories, $g$ can become large in the infrared and the dilute instanton gas approximation can break down. However, we believe that the correct formulation of the axion WGC should apply at an ultraviolet scale such as $f_\theta$, where the coupling $g$ is small. Because small couplings run very slowly, the conclusion is not especially sensitive to the precise choice of ultraviolet scale.

\item {\em A fundamental axion.} In order for the axion strings to lead to interesting conclusions about fundamental high-spin states and a low UV cutoff, we must assume they are fundamental objects, not solitonic strings describable within low-energy effective field theory. Again, the Chern-Weil perspective adds credibility to this assumption. It is clear that a fundamental axion can gauge a $(-1)$-form symmetry; indeed, in examples it generally descends from a gauge field in higher dimensions. It is less clear that this is true of an ordinary pseudo-Goldstone boson. We expect that global symmetries in quantum gravity are at best approximate, and that symmetry-violating operators are present in the effective action. Such operators add terms to the right-hand side of the equation $\partial_\mu j^\mu_\mathrm{PQ} = \frac{1}{8\pi^2} \tr(F \wedge F) + \cdots$, indicating that the current $\tr(F \wedge F)$ is not exact and hence not gauged. In other words, a traditional Peccei-Quinn axion that is not fundamental could leave behind a true global $(-1)$-form symmetry. 

\item {\em The (mild) WGC for $0$-form and $2$-form gauge fields.} Some of the original arguments for the WGC apply to $p$-form gauge fields where $1 \leq p < d-2$, but not to the cases that we have relied on. Nonetheless, in some theories the $0$-form WGC follows from more established cases, e.g., via T-duality~\cite{Brown:2015iha} or dimensional reduction~\cite{Heidenreich:2015nta}. In some theories there is a notion of ``extremal instanton'' that is quite closely analogous to that of extremal black holes~\cite{Bergshoeff:2004fq,Bergshoeff:2004pg}, further motivating the extension of the WGC to axions. It is possible to reformulate the axion WGC in a manner that makes sense when instanton actions are small~\cite{Stout:2020uaf}.  Subtleties related to 4d strings with large decay constants are also discussed in~\cite{Dolan:2017vmn, Hebecker:2017wsu}. The $2$-form WGC has also been directly argued for in terms of black hole physics~\cite{Hebecker:2017uix, Montero:2017yja}, as a necessity for eliminating the multiplicity of axionic black hole states with a nontrivial integral of $B_2$ over the horizon~\cite{Bowick:1988xh}. Thus, the mild $0$-form and $2$-form WGCs are widely accepted, and are perhaps the least controversial of our four major assumptions.

\end{enumerate}

Notice that many of these assumptions may be required by phenomenological aspects of the model, for instance, if we are interested in a model where an axion couples to gauge fields and its potential is dominated by gauge theory instantons. We also see that breakdowns in these assumptions are closely related to alternative mechanisms of eliminating Chern-Weil symmetries. This highlights the importance of further studies of the fate of Chern-Weil symmetries, and especially of $(-1)$-form symmetries, in quantum gravity.

Although simple examples of stringy states obeying the WGC actually feature towers that asymptotically {\em saturate} the WGC bound (e.g.,~\cite{Arkanihamed:2006dz, Heidenreich:2015nta, Heidenreich:2016aqi}), this is not always the case. For example, for gauge fields on D-branes wrapping internal cycles in Type II string theories, axion strings are D-branes wrapping intersecting cycles. When the gauge field cycle is small but the overall volume is large, the strings have tension parametrically below~\eqref{eq:stringtensionbound}. Because charged black holes can (by definition) saturate the WGC bound, it would be interesting to understand how states of large charge interpolate between axion strings and black holes in such examples.

\section{Kaluza-Klein Theory and Axion Couplings}
\label{sec:KKtheory}

\subsection{Abelian Kaluza-Klein Theory and Supersymmetry}
\label{subsec:KKSUSY}

The simplest example of a gauge theory, in a gravitational context, that lacks a $\theta H \wedge H$ term is abelian Kaluza-Klein theory (arising from a circle compactification). (We use the notation $H = \rmd B$ for the Kaluza-Klein gauge field strength, because in this section we will reserve $F$ for the field strength of a different gauge field, to be introduced shortly.) One way to understand the absence of such a coupling is via the Kaluza-Klein monopole~\cite{Gross:1983hb, Sorkin:1983ns}. This monopole cannot be boosted to obtain electric Kaluza-Klein charge, so it has no dyonic degree of freedom. On the other hand, any time an abelian gauge theory admits a $\theta H \wedge H$ term, its magnetic monopoles must have a dyonic collective coordinate for consistency with the Witten effect~\cite{Witten:1979ey}.

At first glance, this might seem puzzling in a supersymmetric context. The gauge coupling in Kaluza-Klein theory is determined by $1/e_\textsc{KK}^2 = \frac{1}{2} R^2 M_{\rm Pl}^2$. The radius $R$ of the compactification is a modulus. In a supersymmetric theory, the coefficients of $H \wedge \star H$ and $H \wedge H$ in the action are related, coming from the real and imaginary part of a holomorphic gauge coupling function. If the gauge coupling appearing in front of $H \wedge \star H$ is a modulus, then the coupling in front of $H \wedge H$ must also be a modulus. How could this be, if there is no $\theta$ term?

This puzzle can be resolved by examining the structure of the 4d $\mathcal{N}=2$ SUSY theory arising from compactifying minimal 5d supergravity. This theory contains not just the Kaluza-Klein gauge field $B$, but another gauge field $A$ descending directly from the 5d graviphoton field $A^{(5)}$, as well as a 4d axion field $\theta$ that arises from the integral of the 5d graviphoton around the circle: 
\begin{align}
\rmd s_5^2 &= \mathrm{e}^{\lambda/2} \rmd s_4^2 + \mathrm{e}^{-\lambda}(\rmd y + R B)^2, &
A^{(5)} &= A + \frac{\theta}{2\pi} \left(\frac{\rmd y}{R} + B\right).
\end{align}
The structure of the gauge theory is not a simple product, but instead has a higher-group structure incorporating the axion $\theta$. In particular, the gauge field $A$ transforms under a $2 \pi$ shift of the axion:
\begin{equation} \label{eq:KKmonodromy}
\theta \mapsto \theta + 2\pi \quad \Rightarrow \quad A \mapsto A - B ,
\end{equation}
and likewise for the field-strength $F = \rmd A$, which thereby fails to be gauge invariant.
Rather, the gauge invariant field strength is
\begin{equation}
\widetilde{F} = F + \frac{\theta}{2\pi} H,
\end{equation}
with the modified Bianchi identity $\rmd {\widetilde F} = \frac{1}{2\pi} H \wedge \rmd \theta$.

The monodromy~\eqref{eq:KKmonodromy} is crucial for making sense of the Chern-Simons terms in the theory. The 5d Chern-Simons term 
\begin{equation} \label{eq:5dCS}
\frac{k}{6 (2\pi)^2} \int A^{(5)} \wedge \rmd A^{(5)} \wedge \rmd A^{(5)}\,,~~~~k \in \mathbb{Z}
\end{equation}
gives rise to 4d terms of the form\footnote{Being total derivatives, the constant shifts in the coefficients of the $F\wedge F$ and $H \wedge F$ terms are not easily determined using classical dimensional reduction. Nonetheless, their presence is required to ensure invariance under large gauge transformations, which is how we have determined their values. In fact, there are multiple distinct solutions consistent with large gauge transformations: $-\pi/3$ can be replaced with $5\pi/3$, and any constant shift can be added to $H\wedge H$ term. A more complete treatment involving anomalies and other subtleties (analogous to, e.g.,~\cite{Witten:1996md}) would determine these shifts unambiguously. We leave this as an interesting direction for future research.}
\begin{equation} \label{eq:thetaterms}
\frac{k}{8 \pi^2} \int \left[(\pi+\theta) F \wedge F +  \left(-\frac{\pi}{3}+\frac{\theta^2}{2 \pi}\right) H \wedge F + \frac{\theta^3}{3 (2 \pi)^2} H \wedge H\right].
\end{equation}
We see that the coefficient of $H \wedge H$ {\em is} dependent on a modulus---but it scales as $\theta^3$, rather than $\theta$ itself! In isolation, this would conflict with the $2\pi$ shift symmetry of $\theta$. However, the monodromy~\eqref{eq:KKmonodromy} and the constant shifts of the $F \wedge F$ and $H \wedge F$ theta terms ensure the consistency of the whole structure.

By supersymmetry, the related $\theta$-terms in~\eqref{eq:thetaterms} are tied up with related moduli-dependence of the gauge couplings themselves. In particular, the gauge coupling of $A$ is given by
\begin{equation}
\frac{1}{e^2} = \left(\frac{\sqrt{3} k}{ 4 \pi e_\textsc{KK}}\right)^{2/3}.
\end{equation}
Because there {\em is} an ordinary $\theta F \wedge F$ coupling, the argument that we gave in \S\ref{sec:argument} applies directly to $A$, implying the existence of axion strings carrying $A$ charge, at or below a mass scale
\begin{equation}
e M_\mathrm{Pl} = \left( \frac{ 4 \pi e_\textsc{KK}}{ \sqrt{3} k } \right)^{1/3} M_\mathrm{Pl}.
\label{eq:third}
\end{equation}
This scale, at which high-spin states must appear, is nothing other than the 5d Planck scale; indeed, this one-third scaling has appeared in past arguments relating WGC towers to emergence~\cite{Heidenreich:2016aqi,Heidenreich:2017sim}. It is also an example of~\cite{Lanza:2021qsu}'s $m^2 \propto T^w$ scaling with $w = 3$. The axion strings of $\theta$ are simply the magnetic monopoles of the graviphoton in 5d, which can carry electric charge via the (5d version of the) Witten effect associated with the Chern-Simons coupling~\eqref{eq:5dCS}. 

The fact that the axion strings are magnetic objects in 5d may seem surprising at first glance, since the WGC allows magnetically charged objects to have tension at or {\em above} the Planck scale. However, at small $e_\textsc{KK}$, the 4d Planck scale is much larger than the 5d Planck scale, so the 5d magnetically charged strings can still have tension {\em below} the 4d Planck scale. For instance, in the case of minimal 5d supergravity, strings charged magnetically under the graviphoton have tension at the 5d Planck scale, and after compactification, they become 4d axion strings with tension at the scale $(eM_\mathrm{Pl})^2$, which is indeed subplanckian. 

\subsection{Kaluza-Klein and Winding U(1)s}
 \label{subsec:KKwinding}

Another case in which the Kaluza-Klein U(1) appears in concert with an additional U(1) is in circle compactifications of string theory, which have a U(1) winding number symmetry. We will continue to denote the field strength of the KK U(1) as $H$, and denote the winding number field strength as $G$. Then these compactifications have a $\theta H \wedge G$ term, where $\theta$ descends from the dual of the string theory $B$-field. Kaluza-Klein monopoles admit winding number charge, and vice versa~\cite{Sen:1997zb, Gregory:1997te}. 

This case serves as both an interesting example and an interesting counterexample to our earlier arguments. First, let us explain how it aligns with our earlier arguments. We can form instantons for $\theta$ via a KK monopole with winding-dyon excitations, or via a winding monopole with KK excitations. The formalism of~\cite{Fan:2021ntg} generalizes to such cases: we expect that the instanton action is 
\begin{equation}
S = \frac{2\pi m_\textsc{M}}{m_\labD},
\end{equation}
where $m_\labD^2/m_\textsc{M}$ is the energy splitting between the dyons and the monopole. As in~\cite{Fischler:1983sc,Fan:2021ntg}, we estimate this by integrating the electric field outside the monopole down to the radius of the monopole core---but now, this electric field is a winding field when the monopole has magnetic KK charge, and vice versa. The calculation leads to the conclusion
\begin{equation}
S \sim \frac{4\pi^2}{e_\textsc{KK} e_\textsc{W}}.
\end{equation}
With this estimate, we repeat our previous arguments, predicting the existence of axion strings at a mass scale
\begin{equation}
M_\mathrm{str} \lesssim \sqrt{e_\textsc{KK} e_\textsc{W}} M_\mathrm{Pl}.
\end{equation}
For Kaluza-Klein and winding modes, this does, in fact, parametrically match the string scale, as $e_\textsc{KK} \sim 1/(R M_\mathrm{Pl})$ and $e_\textsc{W} \sim M_\mathrm{str}^2 R/M_\mathrm{Pl}$.

However, our argument in \S\ref{sec:argument} went a step further, and claimed that anomaly inflow leads us to expect charged modes at the string scale. Here, this is false: for compactifications on a large circle, the winding modes have mass {\em parametrically larger} than $M_\mathrm{str}$, at the scale $M_\mathrm{str}^2 R$. What went wrong with the earlier argument?

The answer  is that in this case, anomaly inflow on the axion string (which is simply an F-string) is satisfied by a mix of left- and right-moving modes,\footnote{Indeed, if we only had left-moving modes carrying KK and winding charge, we would necessarily have $\theta H \wedge H$ and $\theta G \wedge G$ terms as well, by anomaly inflow, and likewise if we only had right-moving modes. To obtain only the mixed $\theta H\wedge G$ term requires a cancellation between the left and right movers.} and the combination of these modes with winding charge cannot be excited at the string scale.
This plays out in a familiar  manner, with left-moving charge $p_L = q_\textsc{KK}/R + q_\textsc{W} R/\alpha'$ and right-moving charge $p_R = q_\textsc{KK}/R - q_\textsc{W}R/\alpha'$ (see, e.g., chapter 8 of~\cite{Polchinski:1998rq}). The operator with minimal winding charge, $q_\textsc{W} = 1$, can only be excited at large values of $p_L$ and $p_R$. Thus, we conclude that in theories with mixed $\theta F_1 \wedge F_2$ terms, axion strings can identify a fundamental cutoff at $\sqrt{e_1 e_2}M_\mathrm{Pl}$, but we can no longer assume that the WGC towers for both gauge groups arise at the same scale. However, we expect that at least {\em one} of the towers will arise at or below the string scale. This is because, as in the case of momentum and winding, there are duality constraints that relate the periods of different worldsheet charged fields, ensuring that one of them can be excited at small energy cost. We discuss these constraints further in Appendix~\ref{app:inflow}.

\subsection{Nonabelian Kaluza-Klein Gauge Groups}

Nonabelian Kaluza-Klein theories also have a tower of charged Kaluza-Klein modes at the Kaluza-Klein scale $g M_\mathrm{Pl}$, which for small $g$ is well below the higher-dimensional Planck scale. Like abelian Kaluza-Klein gauge theories, they lack axion couplings, so the argument of \S\ref{sec:argument} does not apply to these theories. Unfortunately, we do not know of modified arguments for nonabelian Kaluza-Klein theories analogous to those in \S\ref{subsec:KKSUSY} or \S\ref{subsec:KKwinding} that point to the existence of a tower of string modes.

The absence of an axion coupling raises the question of what prevents nonabelian Kaluza-Klein theories from having a global $(-1)$-form Chern-Weil symmetry generated by $\tr(F \wedge F)$. Unfortunately, we lack a general answer. The simplest examples of nonabelian Kaluza-Klein theories are SO$(n+1)$ gauge theories arising from compactifications on $S^n$ and PSU$(n+1)$ gauge theories arising from compactifications on $\mathbb{CP}^n$. These are manifolds of positive curvature, by necessity: compact manifolds with negative-definite Ricci tensor have no Killing vectors (see page 3 of~\cite{Frodden:2020gzi} for a short proof), while the continuous isometries of Ricci-flat manifolds are abelian. Thus, nonabelian Kaluza-Klein theories involve compactifications on manifolds that have positive curvature (at least in some directions). The simplest examples are the Freund-Rubin compactifications~\cite{Freund:1980xh}, which balance the curvature of the internal manifold against the flux of a higher-form gauge field over the internal manifold. As discussed in~\cite{Heidenreich:2020pkc}, the fate of the Chern-Weil symmetry generated by $\tr(F \wedge F)$  in some such cases may be understood via a coupling to a heavy axion that obtains a mass above the Kaluza-Klein scale by coupling to a 4-form flux~\cite{Dvali:2005an}. This suggests that it may be fruitful to  explore how such a large axion mass might modify the WGC  for the axion or its dual $B$-field, or undermine our assumption that the instanton action relevant for the axion WGC is the standard Yang-Mills instanton action.

Another salient feature of Freund-Rubin compactifications is that they have no separation of scales between the AdS curvature radius and the size of the internal dimensions, and so they do not resemble four-dimensional flat space. Thus, they have little phenomenological relevance, and might be less interesting as exceptions to our general picture of axion string towers. On the other hand, there is no obvious obstruction within effective field theory to obtaining a 4d theory with small internal dimensions and nonabelian Kaluza-Klein gauge fields; for example, one could fine-tune terms arising from a higher-dimensional cosmological constant, curvature, and flux to construct a vacuum with zero 4d cosmological constant. Thus, it is unclear whether the absence of a flat-space limit might have any relevance for how we should understand the fate of the Chern-Weil symmetry. At some level, we expect that because the 4d gauge fields in these examples arise directly from higher-dimensional gravity, there should be no new global symmetries that were not already present in the gravitational theory. This would be an interesting topic to explore further; for example, can we find geometries that are higher-dimensional uplifts of Yang-Mills instantons?

Because our understanding of how nonabelian Kaluza-Klein gauge theories fit into our larger picture is so unclear, this area provides an important target for further study.

\section{An Analog in 5d Supergravity}
\label{sec:5dsugra}

So far, we have focused our attention on 4-dimensional axion gauge theories. In that context, we have seen that in the presence of a coupling $\theta F \wedge F$, the axion WGC and the 2-form WGC together imply the existence of a tower of light states charged under the gauge field $F$, in accordance with the Tower and Sublattice WGC.

In higher dimensions, a similar story plays out. In five dimensions, for instance, we may consider a theory of two 1-form gauge fields $A$ and $B$, with a Chern-Simons coupling of the form $B \wedge F \wedge F$, $F= {\rm d} A$. This coupling leads to anomaly inflow on the worldsheet of a fundamental string charged magnetically under $B$, producing a tower of particles charged under $A$.

As in the 4-dimensional case, we would like to say that this tower of light charged particles satisfies the WGC, at least parametrically. However, this relies on a particular scaling $T \sim g_A^2$ of the fundamental string tension $T$ with the gauge coupling $g_A$ of $A$, which followed in 4d from Assumption \ref{number2} in~\S\ref{ssec:Assumptions}. It is not clear that this scaling will hold in 5d in general, but in what follows, we will see that it does hold in any 5d supergravity theory containing exactly two gauge fields $A$, $B$ coupled to each other by a Chern-Simons term $B \wedge F \wedge F$.

We begin with a review of relevant aspects of 5d supergravity, following the conventions of \cite{Alim:2021vhs}. At a generic point on the Coulomb branch, the action for the bosonic fields in a gauge theory with $n$ vector multiplets is given by
\begin{align}
  S &= \frac{1}{2 \kappa_5^2}  \int d^5 x \sqrt{- g}  \left( R -
  \frac{1}{2} \mathfrak{g}_{i j} (\phi) \partial \phi^i \cdot \partial \phi^j
  \right) - \frac{1}{2 g_5^2} \int a_{I J} (\phi) F^I \wedge
  \star F^J \nonumber \\
  &+ \frac{1}{6(2\pi)^2} \int C_{I J K} A^I \wedge F^J \wedge F^K,
  \label{eqn:5dsugra}
\end{align}
where $I = 0, \ldots, n$, $i = 1, \ldots, n$, and $g_5^2 = (2\pi)^{4/3} (2\kappa_5^2)^{1/3}$. The scalar metric
$\mathfrak{g}_{i j} (\phi)$, the gauge kinetic matrix $a_{I J} (\phi)$, and the
Chern-Simons couplings $C_{I J K}$ are all determined by a cubic prepotential
$\mathcal{F} [Y]$, which is a cubic in $Y^I$. The Coulomb branch corresponds to the slice $\mathcal{F} =1$.

Defining $\mathcal{F}_I := \partial_I \mathcal{F}$, $\mathcal{F}_{I J}
:= \partial_I \partial_J \mathcal{F}$ and $\mathcal{F}_{I J K} :=
\partial_I \partial_J \partial_K \mathcal{F}$, the Chern-Simons coupling is given by $C_{I J K} = \mathcal{F}_{I J K}$, and the gauge kinetic matrix is given by
\begin{equation}
  a_{I J} (\phi) =\mathcal{F}_I \mathcal{F}_J 
  -\mathcal{F}_{I J}  \,.
  \label{aIJ}
\end{equation}
It is useful to work in homogenous coordinates, in which we drop the constraint $\mathcal{F}=1$, and instead set 
\begin{equation}
  a_{I J} = \frac{\mathcal{F}_I \mathcal{F}_J }{\mathcal{F}^{4/3}} 
  - \frac{\mathcal{F}_{I J}}{ \mathcal{F}^{1/3} }  \,.
  \label{aIJhom}
\end{equation}

We are interested in the behavior of the gauge kinetic matrix $a_{IJ}$ in an infinite distance limit. As shown in \cite{Heidenreich:2020ptx} and as we will see explicitly in the examples below, such a limit necessarily has a vanishing gauge coupling $g_A \rightarrow 0$. This limit also has a diverging gauge coupling $g_B \rightarrow \infty$, which by the 2-form WGC implies the existence of a magnetic string with tension $T \propto g_B^{-1} \rightarrow 0$ \cite{Alim:2021vhs, BPSstrings}. This string necessarily saturates a BPS bound \cite{BPSstrings}.

For simplicity, we concentrate on the case with $n=2$ gauge fields, as arises from M-theory compactified to 5d on a Calabi-Yau threefold with $h^{1,1}=2$. Consider a point $Y^I_*$ on the Coulomb branch, which lies at infinite distance in moduli space, and a path approaching this point which is linear in homogenous coordinates, i.e., $Y^I = Y^I_* + t Y^I_{(1)}$. By appropriate linear transformations of the coordinates, we may set $Y^I_* = (1, 0)$ and $Y^I_{(1)} = (0, 1)$. The infinite distance limit in question corresponds to $t \rightarrow 0$.

As discussed in \cite{Alim:2021vhs}, the prepotential $\mathcal{F}$ must vanish in the infinite distance limit, hence it takes the general form
\begin{equation}
\mathcal{F} = \beta (Y^0)^2 Y^1 + \gamma Y^0 (Y^1)^2 + \lambda (Y^1)^3 \,.
\end{equation}
There are two cases of interest to us: (a) $\beta = 0$ and (b) $\beta \neq 0$. In case (a), we redefine $Y^0 \rightarrow \gamma^{-1} Y^0 - \lambda Y^1$
to set $\gamma=1$ and $\lambda = 0$, leaving $\mathcal{F} = Y^0 (Y^1)^2$. 
The gauge kinetic matrix is exactly\footnote{Note that the charge lattice need not be integral following the redefinition $Y^0 \rightarrow \gamma^{-1} Y^0 - \lambda Y^1$. Nonetheless, it remains rational, and its exact form will have no impact on our arguments.}
\begin{align}
a_{IJ} = \left( \begin{array}{ccc}
 t^{4/3} & 0 \\
0 & 2 t^{-2/3}  
\end{array} \right) \,.
\end{align}
Therefore, identifying $B=A^0$ and $A = A^1$, we read off $g_B = t^{-2/3} g_5$ and $g_A = t^{1/3} g_5/\sqrt{2}$. 
 We thus have $g_B \sim g_5^3/g_A^2$, so by the 2-form WGC we expect a tensionless string of tension $T \sim \frac{2\pi}{g_B} M_{\textrm{Pl}; 5}^{3/2}$, with string scale $M_s \sim \sqrt{2 \pi T} \sim g_A M_{\textrm{Pl}; 5}^{3/2}$ coinciding with the WGC scale for $A$.\footnote{Here we carefully track $2\pi$'s but not other $O(1)$ factors, as before. Note that $g_5^3 \sim (2 \pi)^2/M_{\textrm{Pl}; 5}^{3/2}$.} 

Next, let us consider case (b), in which $\beta \neq 0$. In this case, we can set $\beta = 1$ by redefining $Y^1 \to \beta^{-1} Y^1$ and then $\gamma = 0$ by shifting $Y^0 \rightarrow Y^0 - \frac{1}{2} \gamma Y^1$. We then have $\mathcal{F} = (Y^0)^2 Y^1 + \lambda (Y^1)^3$. 
At leading order in $t$, the gauge kinetic matrix is
\begin{align}
a_{IJ} = \left( \begin{array}{ccc}
 2 t^{2/3} & 4 \lambda t^{5/3} \\
 4 \lambda t^{5/3} & t^{-4/3} 
\end{array} \right) \,.
\end{align}
Thus, when $\lambda \ne 0$, there is kinetic mixing between $A^0$ and $A^1$. However, this mixing is suppressed in the $t \to 0$ infinite-distance limit. In this limit, identifying $B = A^0$ and $A = A^1$ as before, we obtain $g_B = t^{-1/3} g_5 / \sqrt{2}$ and $g_A = t^{2/3} g_5$, hence $g_B \sim g_5^{3/2} / g_A^{1/2}$.
Thus, by the 2-form WGC we expect a tensionless string of tension $T \sim \frac{2 \pi}{g_B} M_{\textrm{Pl}; 5}^{3/2}$, with string scale $M_s \sim \sqrt{2 \pi T} \sim \sqrt{2 \pi} g_A^{1/4} M_{\textrm{Pl}; 5}^{9/8}$. This one-quarter scaling with the gauge coupling has previously appeared in discussions of WGC towers and emergence~\cite{Heidenreich:2016aqi,Heidenreich:2017sim}; it is the 5d analog of the one-third scaling in 4d that we saw above in \eqref{eq:third}.

What is the salient difference between case (a) and case (b)? In case (a), from the form of the prepotential $\mathcal{F} = Y^0 (Y^1)^2 $, we have $C_{011} \neq 0$, and hence there is a Chern-Simons coupling of the form $B \wedge F \wedge F$. As in 4d, anomaly inflow implies the existence of a tower of charged states on the $B$ string worldsheet, which have string scale mass $M_s \sim g_A M_{\textrm{Pl}; 5}^{3/2}$.

In case (b), on the other hand, the structure of the prepotential gives $C_{011} = 0$, so the Chern-Simons term $B \wedge F \wedge F$ is absent. Instead, there is a $B \wedge F \wedge H$ Chern-Simons coupling.\footnote{There is also a $A \wedge F \wedge F$ coupling when $\lambda \ne 0$, but this has no effect on anomaly inflow on the $B$-string.} By anomaly inflow, this produces both left and right-moving modes on the $B$-string carrying both $A$ and $B$ charge in a manner very much like the momentum/winding examples discussed in~\S\ref{subsec:KKwinding}. As before, this permits the $A$-charged ``momentum'' modes to be excited at a scale $g_A M_{\textrm{Pl}; 5}^{3/2}$ far below the string scale, whereas the $B$-charged ``winding'' modes first appear at a scale $g_B M_{\textrm{Pl}; 5}^{3/2}$ far \emph{above} the string scale; indeed, a quick calculation confirms that $M_s \sim \sqrt{g_A g_B} M_{\textrm{Pl}; 5}^{3/2}$, exactly as expected for momentum and winding modes.

More generally, the Emergent String Conjecture \cite{Lee:2019wij, Lee:2019xtm} suggests that the tensionless string in case (a) should correspond to a type II/heterotic string in some dual frame, whereas case (b) corresponds to a decompactification limit. In this case, the gauge coupling $g_A$ should be identified with the KK gauge coupling $e_{\text{KK}}$, the string scale $M_s \sim g_A^{1/4} M_{\textrm{Pl}; 5}^{9/8}$ is simply the 6-dimensional Planck scale, and the modes discussed above are precisely momentum and winding modes
 of the string on the compact circle.

\section{Points for Further Study}
\label{sec:conclude}

We have given a straightforward argument, for a single axion coupled to a single gauge field, that there should be axion strings with charged excitations at the WGC scale. As the example of Kaluza-Klein and winding U(1) symmetries in \S\ref{subsec:KKwinding} shows, these arguments can be more subtle in the presence of multiple gauge fields. Nonetheless, one of the two towers lies below the string scale. We expect that this generalizes, due to duality constraints on worldsheet fields, as discussed in Appendix~\ref{app:inflow}. This raises the question of whether there is a suitable generalization of our arguments to theories with arbitrarily many axions and gauge fields, together with general Chern-Simons terms. This would be analogous to the Convex Hull Condition for the ordinary WGC~\cite{Cheung:2014vva}, but because the Chern-Simons terms link axion symmetries and ordinary gauge symmetries into a higher group structure, there could be novel complications. Notice that the existence of different axion strings for independent gauge groups does not automatically imply the T/sLWGC, because if the strings cannot form bound states, then there will not be single-particle states occupying the sites in the charge lattice carrying both gauge charges.

Another natural generalization is to higher dimensions. The results of \S\ref{sec:5dsugra} suggest that our argument can be adapted away from 4d. However, the abelian instanton calculation of~\cite{Fan:2021ntg} is not so easily translated to different numbers of dimensions, since the higher-dimensional monopole worldvolume may not give rise to a well-defined semiclassical expansion as in the worldline case.

Kaluza-Klein theory has often been discussed as a paradigmatic example of core Swampland conjectures: there is a tower of charged modes obeying the WGC, their masses become exponentially light as a function of the canonically normalized radion field as the Distance Conjecture requires, and the gauge coupling vanishes in the infinite-distance limit. However, the perspective of this paper suggests that in some ways Kaluza-Klein theory may be an exception to the rule, with theories of low-tension strings providing the more generic example of a weak-coupling limit. Our work has obvious connections to the Emergent String Conjecture of~\cite{Lee:2019wij} and the Distant Axionic String Conjecture of~\cite{Lanza:2020qmt,Lanza:2021qsu}. It would be worthwhile to explore these connections further.

An important question for particle physics is when consistency of the theory requires the existence of a light axion with a $\theta F \wedge F$ coupling. This question is potentially relevant for the Strong CP problem, the nature of dark matter, and the dynamics of inflation. We have now seen that it is also relevant for understanding the nature of the ultraviolet cutoff of a weakly coupled gauge theory. Theories lacking $\theta F \wedge F$ terms exist. Aside from the Kaluza-Klein theories we discussed in \S\ref{sec:KKtheory}, other examples arise from compactifications on rigid Calabi-Yau 3-folds (with $h^{2,1} = 0$)~\cite{Cecotti:2018ufg}. However, these also have gauge couplings frozen to particular values, rather than determined by a light modulus, and so they cannot be tuned to arbitrarily weak coupling. Are Kaluza-Klein theories the only ones with tunably small couplings and no light axion? Are there any examples of quantum gravities containing weakly coupled gauge fields with {\em chiral} matter, but without $\theta F \wedge F$ couplings?

In the theories to which our argument applies, the WGC scale $gM_\mathrm{Pl}$ is a scale at which local QFT irrevocably breaks down and a theory including higher-spin states is necessary. For phenomenological theories with small $g$, this can provide much more stringent constraints than would arise from the $g^{1/3}M_\mathrm{Pl}$ bound discussed in~\cite{Heidenreich:2016aqi, Heidenreich:2017sim}. For example, given that experimental constraints imply that a massless $B-L$ gauge boson must have coupling $g \lesssim 10^{-24}$~\cite{Wagner:2012ui, Heeck:2014zfa}, a gravitational cutoff at $g^{1/3} M_\mathrm{Pl}$ is perfectly compatible with data, whereas a gravitational cutoff at $g M_\mathrm{Pl} \lesssim \mathrm{keV}$ is ruled out. However, this more stringent bound only applies under the assumption that a fundamental axion coupling to the gauge field exists. If the only quantum gravities allowing extremely weakly coupled gauge theories without axion couplings are Kaluza-Klein theories, this would exclude the possibility of a massless $B-L$ gauge boson (because chiral Standard Model fermions carry $B-L$ charge, an impossibility for Kaluza-Klein gauge theories). This provides one example of the phenomenological importance of understanding the space of consistent gauge theories without fundamental axion couplings in quantum gravity. We will leave further exploration of phenomenological applications for future work.

A major emerging theme in recent work has been the importance of higher-group symmetries, which link $p$-form gauge theories of different $p$ into a larger structure via Chern-Simons terms. We should think of the Weak Gravity Conjecture, in its fullest form, as constraining these higher-group gauge symmetries. The consistency among different points of view is noteworthy: by invoking the WGC for the axion and its dual 2-form, and assuming that the relevant instantons are standard gauge theory instantons, we have derived the WGC for a 1-form gauge field (up to an order-one prefactor) as an output.

This work is largely an application of the ideas about Chern-Weil symmetries introduced in~\cite{Heidenreich:2020pkc}. It reinforces the importance of answering questions raised there: how does quantum gravity ensure that all Chern-Weil symmetries are gauged or broken? In particular, when do axions exist to gauge $(-1)$-form instanton number symmetries, and how can these symmetries be broken in the absence of axions? How are they broken in nonabelian Kaluza-Klein theories? Is it possible to consistently gauge an instanton number symmetry with chiral fermions, rather than a fundamental axion? How should we interpret the axion WGC in the presence of axion couplings to 4-form fluxes (as in axion monodromy~\cite{Dvali:2005an,Silverstein:2008sg,McAllister:2008hb,Kaloper:2008fb})?

There is a rich, emerging story linking the Weak Gravity Conjecture, the physics of axions (which is of enormous phenomenological interest), and higher-group symmetries. We are eager to follow it wherever it may lead us.

\section*{Acknowledgments}
MR is supported in part by the DOE Grant DE-SC0013607 and the NASA Grant 80NSSC20K0506. TR is supported by NSF grant  PHY-1820912, the Simons Foundation, and the Berkeley Center for Theoretical Physics. BH is supported by NSF grants PHY-1914934 and PHY-2112800. This work was performed in part at the Aspen Center for Physics, which is supported by the National Science Foundation grant PHY-1607611.

\appendix

\section{Further Comments on Abelian Instantons}
\label{app:monopolecheck}

In the main text, we have asserted that the action of abelian gauge theory ``instantons,'' in the form of monopole loops with winding, scales parametrically as $\sim 8\pi^2/e^2$. In fact, what was computed in~\cite{Fan:2021ntg}, following the calculation of the dyon energy in~\cite{Fischler:1983sc}, was that $S \sim (4\pi^2/e^2) \sqrt{\max(r_c, r_0)/r_c}$. Here $r_c \equiv \pi/(e^2 m_\textsc{M})$ is the classical radius of the magnetic monopole (of mass $m_\textsc{M}$) and $r_0 = e/(8 \pi^2 f_\theta)$ is the screening length of the axion near the monopole. Implicitly, then, we have assumed that $r_c \gtrsim r_0$. What happens in the opposite limit?

Suppose that $r_0 > r_c$. By assumption, then, we have
\begin{equation}
f_\theta < \frac{e^3 m_\textsc{M}}{8\pi^3}.
\end{equation}
Combining this with the WGC for magnetic monopoles,  $m_\textsc{M} \lesssim \frac{2\pi}{e} M_\mathrm{Pl}$, we learn that
\begin{equation}
f_\theta \lesssim \frac{e^2 M_\mathrm{Pl}}{4\pi^2}.
\end{equation}
This is essentially the same bound
as we had previously derived using the mild axion WGC (up to an unimportant factor of $2$). Thus, even in the limit where the instanton action is parametrically different from our assumption, we recover the same conclusion, again using the mild WGC (but this time for magnetic monopoles). This shows that our argument is robust against the subtlety in the calculation in~\cite{Fan:2021ntg}.

\section{Anomaly Inflow Mini-Review}
\label{app:inflow}

In this appendix, we will give a self-contained review of the physics of anomaly inflow and the existence of localized, charge-carrying degrees of freedom on the worldsheets of axion strings. For early work on this topic, see~\cite{Callan:1984sa, Naculich:1987ci}; for a more extensive review, see~\cite{Harvey:2005it}.

\subsection{Axion Electrodynamics and Magnetic Field Strengths}

Consider a single axion $\theta$ with field strength $G = \rmd \theta$ and a set of U(1) gauge fields $A^I$ with field strengths $F^I = \rmd A^I$, with action
\begin{equation} \label{eq:actionorig}
S = \int \left[ -\frac{1}{2} f_\theta^2 G \wedge \star G - \frac{1}{2} K_{IJ} F^I \wedge \star F^J + \frac{1}{8\pi^2} k_{IJ} \theta F^I \wedge F^J \right].
\end{equation}
Here $k_{IJ}$ is a symmetric matrix of integer Chern-Simons levels, and $K_{IJ}$ is a symmetric kinetic matrix determining the strength of the gauge interactions. We work in conventions where the quantization of fluxes is
\begin{equation}
\frac{1}{2\pi} \oint F^J, \quad \frac{1}{2\pi} \oint G, \quad K_{IJ} \oint \star F^J,\quad  f_\theta^2 \oint \star G \quad \in \mathbb{Z},
\end{equation}
where $\oint$ always denotes an integral over a closed manifold of the appropriate dimension.
The action~\eqref{eq:actionorig} leads to equations of motion
\begin{align} \label{eq:EOMorig}
K_{IJ} \rmd \star F^J &= \frac{1}{4\pi^2} k_{IJ} G \wedge F^J, \nonumber \\
f_\theta^2 \rmd\star G &= -\frac{1}{8\pi^2} k_{IJ} F^I \wedge F^J,
\end{align}
which may be viewed as modified Bianchi identities for magnetic dual gauge field strengths.

We define the dual gauge field strengths (2-forms ${\tilde F}_I$ and a 3-form ${\tilde H}$):\begin{equation} \label{eq:dualityconstraints}
\frac{1}{2\pi} {\tilde F}_I = K_{IJ} \star F^J, \quad \frac{1}{2\pi} {\tilde H} = f_\theta^2 \star G.
\end{equation}
We introduce 1-form magnetic gauge fields ${\tilde A}_I$  and a 2-form magnetic gauge field $\tilde B$ to reproduce the modified Bianchi identities inferred from~\eqref{eq:dualityconstraints} and~\eqref{eq:EOMorig}:
\begin{align} \label{eq:magfields1}
{\tilde F}_I &= \rmd {\tilde A}_I + \frac{1}{2\pi} k_{IJ} \theta F^J + (\text{localized}), \nonumber \\
{\tilde H} &= \rmd {\tilde B} - \frac{1}{4\pi} k_{IJ} A^I \wedge F^J + (\text{localized}).
\end{align}
Here ``localized'' stands in for terms living on the worldvolumes of magnetically charged objects, which will become necessary if we wish to discuss magnetic monopoles and axion strings simultaneously. For our purposes, however, we can focus on axion strings which are electrically charged under $\tilde B$ (magnetically charged under $\theta$). The terms proportional to Chern-Simons levels $k_{IJ}$ encode the familiar physics of modified Bianchi identities and the Witten effect. In particular, the field strengths in~\eqref{eq:magfields1} are gauge invariant because the magnetic gauge fields transform under ordinary electric gauge transformations:
\begin{alignat}{6} \label{eq:modifiedgauge}
\tilde B &\mapsto \tilde B + \frac{1}{4\pi} k_{IJ} \lambda^I F^J & \quad \text{when} \quad & A^I &&\mapsto A^I + \rmd \lambda^I, \nonumber \\
\tilde A_I &\mapsto \tilde A_I - k_{IJ} A^J & \quad \text{when} \quad & \theta &&\mapsto \theta + 2\pi.
\end{alignat}
From this, it immediately follows that we cannot simply define couplings on an axion string worldvolume $\Sigma$ or a magnetic monopole worldline $\Gamma^I$ of the form $\int_\Sigma {\tilde B}$ or $\int_{\Gamma^I} {\tilde A}_I$. These are not, on their own, gauge invariant. This is the essence of anomaly inflow: it requires the presence of localized terms on the worldvolume of magnetically charged objects, which restore gauge invariance in the presence of the modified transformation laws~\eqref{eq:modifiedgauge} induced by Chern-Simons terms. Another way to state the anomaly inflow requirement is to notice that the equations of motion~\eqref{eq:EOMorig} derived from~\eqref{eq:actionorig} are inconsistent in the presence of magnetic sources, because when $\rmd G \neq 0$ and $\rmd F^I \neq 0$ we find that $\rmd(\rmd {\tilde F}_I) \neq 0$ and $\rmd(\rmd {\tilde H}) \neq 0$. We will now see how to resolve this difficulty in the case of axion strings. In the case of magnetic monopoles, an analogous argument points to the existence of the dyon collective coordinate that implements the Witten effect.

\subsection{Localized Modes on Axion Strings}

Axion strings carry charge under $\tilde B$, so the best starting point is an action defined in terms of the $A^I$ and $\tilde B$:
\begin{equation} \label{eq:bulkactionaxion}
S = \int \left[- \frac{1}{2} K_{IJ} F^I \wedge \star F^J - \frac{1}{2 \eB^2} {\tilde H} \wedge \star {\tilde H} \right],
\end{equation}
with $\tilde H$ defined as in \eqref{eq:magfields1}. For the remainder of this discussion, then, we can think of the axion field strength as a derived quantity, $\frac{1}{2\pi} G = \frac{1}{\eB^2} \star {\tilde H}$. The bulk equations of motion derived from~\eqref{eq:bulkactionaxion} are equivalent to those derived from~\eqref{eq:actionorig}.

In the presence of an axion string with worldsheet $\Sigma$, one way to see the need for localized degrees of freedom is to note that we have a contradiction when taking the exterior derivative of the gauge field equation of motion:
\begin{equation}
0 = \frac{1}{2\pi} \rmd( \rmd {\tilde F}_I) = K_{IJ} \rmd(\rmd \star F^J) \overset{?}{=} \frac{1}{4\pi^2} k_{IJ} \rmd G \wedge F^J = \frac{1}{2\pi} k_{IJ} \delta(\Sigma) \wedge F^J \neq 0.
\end{equation}
We can resolve this problem if the $\tilde F_I$ Bianchi identity includes localized degrees of freedom. Start with the ansatz 
\begin{equation} \label{eq:Bianchiansatz}
\frac{1}{2\pi} \rmd {\tilde F}_I = \frac{1}{4\pi^2} k_{IJ} G \wedge F^J + \xi_I \wedge \delta(\Sigma).
\end{equation}
This is consistent provided that
\begin{equation}
\rmd  \xi_I = -\frac{1}{2\pi } k_{IJ} F^J \quad \Rightarrow \quad \xi_I =  -\frac{1}{2\pi} k_{IJ} A^J + \rmd \eta_I.
\end{equation}
Here $\eta_I$ is some worldsheet operator which carries gauge charge; in particular, $\eta_I \mapsto \eta_I + \frac{1}{2\pi} k_{IJ} \lambda^J$ when $A^I \mapsto A^I + \rmd \lambda^I$.
There are potentially different consistent worldsheet theories in which these charged degrees of freedom take different forms. 

Rather than discussing the most general possibility, we will attempt to construct a minimal example of a consistent worldsheet theory. We introduce worldsheet-localized periodic scalar fields $\sigma^I \cong \sigma^I + 2\pi$ (one for each gauge field), which shift under a gauge transformation of $A^I$:
\begin{equation}
\xi_I = -\frac{1}{2\pi} k_{IJ}(A^J + \rmd \sigma^J) \equiv -\frac{1}{2\pi} k_{IJ} \rmd_A \sigma^J.
\end{equation}
With this definition, we have a Bianchi identity
\begin{equation} \label{eq:sigmaBianchi}
\rmd(\rmd_A \sigma^I) = \rmd(\rmd \sigma^I + A^I) = F^I.
\end{equation}
Combined with the earlier~\eqref{eq:modifiedgauge}, we now have  the gauge transformation structure
\begin{equation}
A^I \mapsto A^I + \rmd \lambda^I, \quad \sigma^I \mapsto \sigma^I - \lambda^I, \quad \tilde B \mapsto \tilde B + \frac{1}{4\pi} k_{IJ} \lambda^I F^J.
\end{equation}
The next task is to exploit this structure to build a consistent coupling of the $\tilde B$ field to the axion string.  A worldvolume theory of the form
\begin{equation} \label{eq:stringaction}
\int_\Sigma \left[-\tilde B - \frac{1}{4\pi} k_{IJ} \sigma^I F^J - \frac{1}{4} \beta_{IJ} \rmd_A \sigma^I  \wedge \star \rmd_A \sigma^J\right]
\end{equation}
suffices. The negative sign of the $\tilde B$ term follows from the convention $\rmd G = 2\pi \delta(\Sigma)$ that we have chosen. The last term parametrizes the kinetic term  for the localized scalar fields; the $\star$ acting on $\rmd_A \sigma^J$ should be understood as the worldsheet Hodge star. The factor of $1/4$ (rather than $1/2$) in front of the kinetic term is chosen because, as we will see momentarily, this is really a pseudo-action that we will supplement with duality constraints, rather than a proper action.

Varying with respect to $\sigma$ and using~\eqref{eq:sigmaBianchi}, we find
\begin{equation}
\frac{1}{2} \beta_{IJ} \rmd \star \rmd_A \sigma^J = \frac{1}{4\pi} k_{IJ} F^J = \frac{1}{4\pi} k_{IJ} \rmd(\rmd_A \sigma^J).
\end{equation}
This equation allows us to consistently impose a self-duality constraint, 
\begin{equation} \label{eq:selfduality}
\beta_{IJ} \star \rmd_A \sigma^J = \frac{1}{2\pi} k_{IJ} \rmd_A \sigma^J.
\end{equation}
We impose this constraint because we seek a {\em minimal} way to match the anomaly, introducing  the smallest number of new degrees of freedom. The constraint does not follow from the variation of~\eqref{eq:stringaction}, but is an additional equation that supplements the equations of motion. This indicates that~\eqref{eq:stringaction} is a democratic pseudo-action.\footnote{See~\cite{Bergshoeff:2001pv} for early work on such democratic formulations, and Appendix A of~\cite{Heidenreich:2020upe} for a recent review.} The self-duality constraint makes sense when $k_{IJ}$ is a unimodular matrix: $\beta_{IJ} \star \rmd_A \sigma^J$ and $\frac{1}{2\pi} \rmd_A \sigma^I$ are both integrally quantized.\footnote{Naively, the invariance of $\mathrm{e}^{iS}$ under large gauge transformations for the action~\eqref{eq:stringaction} seems to imply a stronger constraint: the bilinear form $k_{I J}$ should be \emph{even}. However, there are ways to avoid this constraint, e.g., using the Green-Schwarz mechanism, so it does not apply universally.}

As a consistency check, we can compute the equation of motion of the gauge field by varying the combined bulk action~\eqref{eq:bulkactionaxion} and string pseudo-action~\eqref{eq:stringaction}:\footnote{The first two terms on the right-hand side of the first line of~\eqref{eq:eomfull} come from the variation of the ${\tilde H} \wedge \star {\tilde H}$ term in the bulk action. In particular, this gives rise to a localized term on $\Sigma$ because $\delta {\tilde H}$ contains a $\rmd(\delta A^I \wedge A^J)$ term, and $\star \tilde H \sim G$, so we obtain a term $\sim \int [\rmd(\delta A^I \wedge A^J) \wedge G] \sim -\int_\Sigma \delta A^I \wedge A^J$ in the variation of the bulk action after integrating by parts and applying the nontrivial Bianchi identity for $G$. The terms in brackets on the right-hand side of the first line come from varying the string pseudo-action~\eqref{eq:stringaction}. Notice that the localized contributions from {\em both} the bulk and string actions are necessary to arrange the final result in terms of the manifestly gauge-invariant quantity $\rmd_A \sigma^J$, as on the second line.}
\begin{align} \label{eq:eomfull}
K_{IJ} \rmd \star F^J &= \frac{1}{4\pi^2} k_{IJ} G \wedge F^J - \frac{1}{4\pi} k_{IJ} A^J \wedge \delta(\Sigma) + \left[-\frac{1}{4\pi} k_{IJ} \rmd \sigma^J - \frac{1}{2} \beta_{IJ}\star \rmd_A \sigma^J\right] \wedge \delta(\Sigma) \nonumber \\
&= \frac{1}{4\pi^2} k_{IJ} G \wedge F^J - \left[\frac{1}{4\pi} k_{IJ} \rmd_A \sigma^J + \frac{1}{2} \beta_{IJ}\star \rmd_A \sigma^J\right] \wedge \delta(\Sigma).
\end{align}
This reproduces our ansatz for the Bianchi identity~\eqref{eq:Bianchiansatz}, after invoking~\eqref{eq:selfduality}.

Let us comment on two special cases:

First, if we have a single gauge field and $k = 1$, then we introduce a single scalar $\sigma$ obeying the constraint $\beta \star \rmd_A \sigma = \frac{1}{2\pi} \rmd_A \sigma$. From this, it follows that $\beta = \frac{1}{2\pi}$, so we have no freedom to tune the energy associated with excitations of $\sigma$. This case corresponds to the chiral boson, which is related by bosonization to a charged chiral fermion.\footnote{The original work on bosonization is~\cite{Coleman:1974bu,Mandelstam:1975hb}; see~\cite{Karch:2019lnn} for a recent discussion.} This is the familiar minimal case of anomaly inflow, with a massless chiral charged mode that can be excited to produce a charged state with string-scale mass.

Second, if we have two gauge fields with $k = \begin{pmatrix} 0 & 1 \\ 1 & 0 \end{pmatrix}$, then we introduce two fields $\sigma^1$ and $\sigma^2$. Let us further assume that the matrix $\beta$ is diagonal, $\beta = \begin{pmatrix} \beta_1 & 0 \\ 0 & \beta_2 \end{pmatrix}$. In this case, the duality constraint implies that $\beta_1 \beta_2 = \frac{1}{4\pi^2}$. We can think of one of the two scalars as an ``electric'' field and the other as a ``magnetic'' field, and their couplings are inversely related via Dirac quantization. This is precisely the case that arises for Kaluza-Klein and winding symmetries. Ordinarily, we work in terms of a field $X = \sqrt{\beta_1} \sigma^1$ or its $T$-dual coordinate $X' = \sqrt{\beta_2} \sigma^2$; here, we have instead constructed a democratic formulation with both fields appearing in a single pseudo-action. Because the coefficients of the kinetic terms are inversely related, one set of charged modes (conventionally, the Kaluza-Klein modes) can be excited at energy well below the string scale while the other (conventionally, the winding modes) have energy well above the string scale.

These special cases illustrate the general pattern: we cannot claim that there are charged modes below the string scale for any gauge group interacting with an axion, since winding number is a counterexample. However, in such cases the self-duality constraints will dictate that there are other charged modes below the string scale, and the relative mass scales of the different charged states will be related via the duality constraints.\footnote{A different argument for the necessity of  light charged modes below the scale of an axion string's tension is in~\cite{Brennan:2020ehu}.}

\bibliography{ref}
\bibliographystyle{utphys}

\end{document}